# Runge-Lenz operator in momentum space

S. P. Efimov[1]

Bauman Moscow State Technical University, Moscow,105005, Russian Federation

(Dated: November 20, 2024)

## Abstract

The fundamental quantum Coulomb problem in the momentum space is considered. Adifferential equation with SO(4) simmetry has been obtained in the momentum space instead of the integral Fock equation. The corresponding equation. in the coordinate space is the sum of the squares of the angular momentum and the Runge-Lenz operators.This approach is unknown in the momentum space where the Runge-Lenz operator is not applied in the existing theory. The Runge-Lenz operator obtained in the momentum space is simplier than that in the coordinate space and allows one to effectively consider the Coulomb problem in the momentum space. A relation of new operator to the infinitesimal rotation operator of the three-dimensional Fock's sphere has been determined.

## I.  INTRODUCTION

The fundamental quantum Coulomb problem wich allows one to calculate spectrum of a system of two opposite charges, is still relevant in the quantum theory [1-4]. Such founders of twentieth century physics as N. Bohr, A. Sommerfeld, W. Pauli, E. Shrödinger and V. Fock contribute to it. The introduction to the theory of atomic spectra begins with it, and it is thoroughly studied using the theory of special functions. Due to its simplicity and underlying SO(4) group of rotations about a fixed point in four-dimensional (4D) Euclidean space, this problem is an extremely useful and fine tool of theoretical physics for constucting various consepts [5-7].

The transition from the coordinate space to the momentum space is exclusevely efficient in theoretical physics particularly in the quantum electrodynamics, as far as local differential operators transformed to polynomials and considered transformations are reduced to algebraic. The Coulomb problem is in this case of particular significance.  The Shrödinger equation (**SE**) in the momentum space becomes an integral equation. Fock applied stereographic projection to the three-dimensional (3D) momentum space mapping it to the 3D sphere embeded in the 4D momentum space [8-10]. The integral Fock equation is transformed to the equation for spherical functions on 3D sphere in the 4D space, which can be interpreted conditionally as free motion of a particle on the 3D sphere. It reminds motion of a physical point mass in oscillator to and fro when the corresponding point in phase space has free motion on a circle.

---

[1] e-mail: serg.efimo2012@yandex.ru

Let us recall the background proceeding Fock's accomplishment. Two classical vector integrals, the angular momentum and the Laplace-Runge-Lenz vector, in quantum mechanics correspond to vector operators that commute with the energy operator, i.e. with the Hamiltonian. An analysis of their commutators in [11] shows that they generate a Lie algebra (a linear space with a commutation operation) coinsiding with the Lie algebra of infinitesimal rotations in 4D space [1, 3].

For physicists, this correspondence means that some transformation of variables and operators transforms the original quantum Coulomb problem to the state of a particle laid on 3D sphere embedded in 4D momentum space. The energy operator is then invariant under rotations of the 3D sphere, thus a vector creation operator arises naturally which was developed for 2D sphere in [13].

In [14], the Schrödinger equation (**SE**) was transformed so that radii of all orbits are reduced to unity. This means that the problem is reduced to the quantization of charge $Z = n$. The **SE** is squared here, the resulting operator is no longer hermitian but holds all the necessary physical properties. After that, the transition to the momentum space with the locality property can be performed. As a result, we come to a differential equation for eigenfunctions in the momentum space. We find physical meaning of differential equation, determine eigenfunctions and obtain the Runge-Lenz operator of simple differential form. The application of the Runge-Lenz operator is thus simplified considerably. The relation between this operator and the infinitesimal rotation operator of the 4D Fock sphere is found as well.

Consequently, the quantum Coulomb problem in the momentum space is adequate to that in the coordinate space and is very useful for theories including the Coulomb interaction and perturbation theory together.

## II. THE FOCK THEORY

In atomic units, where $\frac{Z^2 m e^2}{\hbar^2}$ is the energy unit and the Bohr radius $a_B = \frac{\hbar^2}{Z m e^2}$ is the length unit, the Schrödinger equation (**SE**) for eigenfunctions has the form

$$\left(-\frac{1}{2}\Delta - \frac{1}{r}\right)\Psi_{nl} = -\frac{1}{2n^2}\Psi_{nl}. \tag{1}$$

Further, it is convenient to convert each orbit radius $na_B$ to single radius [1], i.e., to substitute the position-vector $\mathbf{x}' = \frac{\mathbf{x}}{n}$ for each eigenfunction. Thus, Eq. (1) takes deceptively simple form,

$$(-\Delta + 1)\Psi_{nl} = \frac{2n}{r}\Psi_{nl}, \tag{2}$$

where **x** and $r$ are again the position vector and its magnitude respectively. Then, eigenfunctions in the momentum representation have the scaled argument $\mathbf{p}' = n\mathbf{p}$.

For the transition to the momentum space, eigenfunctions of Eq. (2) to be represented by the Fourier transform ($\hbar = 1$) as

$$\Psi_{nl}(\mathbf{x}) = \frac{1}{(2\pi)^3} \int a_{nl}(\mathbf{p}) e^{i(\mathbf{p}\mathbf{x})} d^3\mathbf{p}. \tag{3}$$

Since the potential $\frac{1}{r}$ is transformed to $\frac{4\pi}{p^2}$, the **SE** becomes non-local and has the form

$$(p^2 + 1) a_{nl}(\mathbf{p}) - \frac{2n}{2\pi^2} \int \frac{a_{nl}(\mathbf{p}') d^3\mathbf{p}'}{|\mathbf{p} - \mathbf{p}'|^2} = 0. \tag{4}$$

Fock applied the stereographic projection [11], which maps 3D plane into a 3D sphere embedded in 4D momentum space, to this equation. The relations of the coordinates of the 4D space on the 3D sphere to the momentum and the constraint on these coordinate have the form[2]:

$$\boldsymbol{\xi} = \frac{2\mathbf{p}}{(1+p^2)}, \quad \xi_0 = \frac{(p^2-1)}{(p^2+1)}, \quad \boldsymbol{\xi}^2 + \xi_0^2 = 1. \tag{5}$$

In the new variables, taking into account the factor chosen by Fock for the function $a_{nl}(\mathbf{p})$, the eigenfunction becomes

$$b_{nl}(\boldsymbol{\xi}, \xi_0) = (p^2 + 1)^2 a_{nl}(\mathbf{p}). \tag{6}$$

It is essential here that the projection is a conformal map. In this case, angles between intersecting curves are conserved. Just for this reason namely, the stereographic projection was invented for nautical charts. The metric on the sphere is expressed in terms of the coordinates of 3D **p**-plane as

$$\frac{4}{(p^2+1)^2} (d\mathbf{p}^2). \tag{7}$$

Hence, the contraction coefficient for elements of the momentum space is $(1+\mathbf{p}^2)/2$. The volume element in Eq. (4) is expressed in terms of the 3D surface element as:

$$d^3p = \frac{1}{8}(1+\mathbf{p}^2)^3 dS_3. \tag{8}$$

The kernel of the integral can be (very fortunately but not obviously) transformed as

$$\frac{1}{|\mathbf{p} - \mathbf{p}'|^2} = \frac{2}{(p^2+1)[(\boldsymbol{\xi} - \boldsymbol{\xi}')^2 + (\xi_0 - \xi_0')^2](p'^2+1)} \tag{9}$$

which does not follow from the conformal property.

Now, substituting Eqs. (6), (8) and (9) into integral equation (4), we obtain

---
[2] In monography [3] the sign of $\zeta$ changed.

$$b_{nl}(\xi, \xi_0) - \frac{n}{2\pi^2} \int \frac{b_{nl}(\xi', \xi_0) dS_3'}{[(\xi - \xi')^2 + (\xi_0 - \xi_{0'})^2]} = 0. \tag{10}$$

As Fock noted, it is the equation for spherical functions on 3D spere [15].

Solutions needed in physics are proportional to (ordinary) 2D spherical functions:
$$Y_{n,l}(\theta, \phi) P_{n-1-l}^{l+1}(\xi_0), \tag{11}$$
where the second factor is the Gegenbauer polynomial [15]. The polynomial argument $\xi_0$ is the forth coordinate on the Fock sphere by Eq.(5). Therefore, Fock found for the first time the general formula for eigenfunctions in the momentum space.

### III. DIFFERENTIAL FORM OF SE IN MOMENTUM SPACE

#### A. Derivation

Following [13], we multiply Eq.(2), where the radius of the orbit (when multiplied by $n$) is reduced to unity, by scalar $r$ and square both sides:

$$r(-\Delta + 1)r(-\Delta + 1)\Psi_{nl} = 4n^2 \Psi_{nl}. \tag{12}$$

Hence,
$$[r^2(\Delta - 1)^2 + 2(\hat{l}_r + 1)(\Delta - 1)]\Psi_{nl} = 4n^2 \Psi_{nl}, \tag{13}$$
where $\hat{l}_r = (r\nabla)$ is the "degree operator ". According to the Euler theorem, which is valid for each homogeneous function (e.g., for $\frac{1}{r}$ and its degrees) rather than only for polynomials, the operator multiplies a homogeneous polynomial by its degree.

We pass to the new function
$$\Phi_{nl} = (\Delta - 1)^2 \Psi_{nl},$$
which corresponds to multiplying the spectrum by $(p^2 + 1)^2$ (see the Fock method in Eq.(6)). For this, we apply the operator $(\Delta - 1)^2$ to Eq.(13) from the left and swap the operators $\hat{l}_x$ and $\Delta$. As a result, we obtain the following equation for the function $\Phi_{nl}(x)$,

$$[(\Delta - 1)^2 r^2 + 2(\Delta - 1)(\hat{l}_r + 3)]\Phi_{nl} = 4(n^2 - 1)\Phi_{nl}.$$

We can now pass to the spectra $a_{nl}(\mathbf{p})$ and $b_{nl}(\mathbf{p})$ by substituting $\nabla_{\mathbf{x}}$ for $i\mathbf{p}$ and $\mathbf{x}$ for $i\nabla_{\mathbf{p}}$:

$$\left[ -\frac{(p^2 + 1)^2}{4} \Delta_{\mathbf{p}} + \frac{(p^2 + 1)}{2} \hat{l}_{\mathbf{p}} \right] b_{nl} = (n^2 - 1) b_{nl}, \tag{14}$$

$$b_{nl}(\mathbf{p}) = (p^2 + 1)^2 a_{nl}(\mathbf{p}),$$

where operator $\hat{l}_{\mathbf{p}} = (\mathbf{p}\nabla_{\mathbf{p}})$ is the operator multipling any polynomial by its degree. Instead of integral Eq.(10), we obtain the equation that is no more complicated than the **SE** in the coordinate space because it is similar to the **SE** with the sum of two potentials. We have now to reveal its symmetry properties.

## B. Solution.

We look for a solution of the equation in the product form

$$b_{nl}(\mathbf{p}) = Y_l(\mathbf{p}) \frac{1}{(p^2+1)^l} P_k\left(\frac{1}{(p^2+1)}\right), \qquad (15)$$

where $Y_l(\mathbf{p})$ is a solid spherical harmonic (which is a homogeneous polynomial) and $P_k(u)$ is a degree-$k$ polynomial. We use the following properties valid for solid spherical functions:

$$\Delta Y_l(\mathbf{p}) = 0, \; Y_l(c\mathbf{p}) = c^l Y_l(\mathbf{p}), \; (\mathbf{p}\nabla) Y_l(\mathbf{p}) = l Y_l(\mathbf{p}). \qquad (16)$$

The solution of the Eq. (14) in the form of Eq.(15) is a polynomial $P_k(u)$ coinciding with the Gauss function[3]

$$F(\alpha, \beta, \gamma, u) = 1 + \frac{\alpha\beta}{\gamma}\frac{u}{1!} + \frac{\alpha(\alpha+1)\beta(\beta+1)}{\gamma(\gamma+1)}\frac{u}{2!} + \ldots,$$

where the parameters are

$$\alpha = -k, \; \beta = 2l + k + 2, \; \gamma = l + \frac{3}{2},$$

$$k = n - l - 1, \; u = \frac{1}{(p^2+1)}. \qquad (17)$$

We recall that transition to the real spectrum requires multiplying the argument p by $n$.

The found system of solutions (15) exactly corresponds to the Fourier transforms of solutions of the **SE.** This follows from coincidence of their angulur dependances (the orbital quantum number $l$, as well as the angular momentum operator, are the same in the coordinate and momentum spaces). In addition, according to the theory of special functions [15], the coordinate substitution (5) transforms the Gauss function into the Gegenbauer polynomial, i.e., into the Fock solution (11).

## C. Examples

(a) $n = 2$, $l = 0$, $k = 1$ (isotropic state). We double the momentum, returning to radius 2:

$$a_{20}(\mathbf{p}) = (\text{const}) \frac{1}{(1+4\mathbf{p}^2)^2}\left(1 - \frac{2}{(1+4\mathbf{p}^2)}\right).$$

The derivation of this formula using the Hankel transforms is rather cumbersome.

(b) $n = l + 1$, $k = 0$ ($l$ is maximum).

$$\Psi_{n(n-1)}(\mathbf{x}) = Y_{(n-1)}(\mathbf{x}) e^{-r},$$

$$a_{nl}(\mathbf{p}) = \frac{Y_l(\mathbf{p})}{(1+\mathbf{p}^2)^{(2+l)}}.$$

---

[3] Hypergeometric function $_2F_1$.

Returning to the physical argument $n\mathbf{p}$ and using the homogeneity of the polynomial $Y_l(\mathbf{p})$, we obtain

$$a_{nl}(\mathbf{p}) = (\text{const}) \frac{Y_{(n-1)}(\mathbf{p})}{(1+n^2\mathbf{p}^2)^{(n+1)}}.$$

## IV. THE RUNGE-LENZ OPERATOR IN THE MOMENTUM SPACE

The Runge-Lenz operator $\hat{R}_r$ in the dimensionless coordinates is equal to [1]

$$\hat{R}_\mathbf{r} = \frac{\mathbf{r}}{r} - \frac{1}{2}([\mathbf{p}\hat{L}] - [\hat{L}\mathbf{p}]).$$

Passing to unit radius, we have to multiply the operator given by Eq.(18) by $n$ for the commutation rules to correspond to the SO(4) group [1, 11]. Calculating the vector products in Eq.(18), we obtain normalized operator

$$\hat{A}_\mathbf{r} = \frac{n\mathbf{r}}{r} - \frac{1}{2}([\mathbf{p}L] - [L\mathbf{p}]) = \frac{n\mathbf{r}}{r} + \mathbf{r}\Delta - (\hat{l}_\mathbf{r} + 1)\nabla, \tag{19}$$

where $\hat{l}_\mathbf{r} = (\mathbf{r}\nabla)$.

The transition to the momentum space gives an integral operator because it involves the potential $\frac{1}{r}$. This is why the Runge-Lenz operator historically was not introduced [1, 3]. We use the method that allowed me in [13] to obtain a differential equation in the momentum space. The substitution of $\frac{n\mathbf{r}}{r}$ expressed from Eq.(2) into Eq.(19) gives

$$\hat{A}_\mathbf{r} = \mathbf{r}\frac{(\Delta + 1)}{2} - (\hat{l}_\mathbf{r} + 1)\nabla, \tag{20}$$

which allows to apply the Fourier transform effectevely. The Fourier transform of Eq.(20) with substitutions

$$r \to i\nabla_\mathbf{p}, \ \nabla_\mathbf{r} \to i p, \ \hat{l}_\mathbf{r} \to -(\hat{l}_\mathbf{p} + 3) \tag{21}$$

yields the modified the Runge-Lenz operator in the momentum space:

$$\hat{A}_\mathbf{p} = i(\hat{l}_\mathbf{p} + 1)\mathbf{p} - \frac{(\mathbf{p}^2 - 1)}{2}i\nabla_\mathbf{p}, \tag{22}$$

acting on functions $a_{nl}(\mathbf{p})$. To compare with the Fock theory, we have to pass to the space of functions $b_{nl}(\mathbf{p})$ from Eq.(14):

$$\hat{A}_\mathbf{p} = (p^2 + 1)^2 i \left[(\hat{l}_\mathbf{p} + 1)\mathbf{p} - \frac{(\mathbf{p}^2 - 1)}{2}\nabla_\mathbf{p}\right]\frac{1}{(p^2 + 1)^2} = i\mathbf{p}\hat{l}_\mathbf{p} - \frac{(\mathbf{p}^2 - 1)}{2}i\nabla_\mathbf{p}. \tag{23}$$

## V. CHARACTERISTICS OF THE RUNGE-LENZ OPERATOR

The Runge-Lenz operator is simpler than that in the coordinate space:

$$\hat{A}_{\mathbf{p}} = i\mathbf{p}\hat{l}_{\mathbf{p}} - \frac{(\mathbf{p}^2-1)}{2}i\nabla_{\mathbf{p}}. \qquad (24)$$

Its characteristics in momentum space are conserved. It has commutation properties:

$$\{L_i, \hat{A}_{\mathbf{p}k}\} = ie_{ikl}\hat{A}_{\mathbf{p}l}, \; \{\hat{A}_{\mathbf{p}i}, \hat{A}_{\mathbf{p}k}\} = ie_{ikl}L_l, \qquad (25)$$

and orthogonality property

$$\hat{A}_{\mathbf{p}}\hat{L} = \hat{L}\hat{A}_{\mathbf{p}} = 0.$$

At last, the most important for the SO(4) group, the sum of squared operators is:

$$\hat{L}^2 + \hat{A}_{\mathbf{p}}^2 = -\frac{(p^2+1)^2}{4}\Delta_{\mathbf{p}} + \frac{(p^2+1)}{2}\hat{l}_{\mathbf{p}}. \qquad (26)$$

The comparison with the operator in the square brackets in Eq. (14) shows that the eigenvalue of operator given by Eq. (26) is equal to $(n^2-1)$; i.e., it is the same as in the coordinate space [1].

We need to testify now which operator corresponds to found the Runge-Lenz operator. According to Eq.(5), an arbitrary function $f(\xi,\zeta)$ on the Fock sphere is transformed after the transition to the 3D momentum plane (i.e., volume) to the function

$$f\left(\frac{2\mathbf{p}}{(p^2+1)}, \frac{(p^2-1)}{(p^2+1)}\right), \qquad (27)$$

where the sum of squared arguments is equal to unity. We apply to it the Runge-Lenz operator (24). Successive differentiation of these arguments gives the following operators:

$$\left(\hat{l}_{\mathbf{p}}\frac{2\mathbf{p}}{(p^2+1)}\right)\nabla_{\xi} = \frac{2\mathbf{p}(1-p^2)}{(p^2+1)^2}\nabla_{\xi} = (-\zeta)(\xi\nabla_{\xi}),$$

$$\left(\hat{l}_{\mathbf{p}}\frac{(p^2-1)}{(p^2+1)}\right)\frac{\partial}{\partial\zeta} = \frac{4p^2}{(p^2+1)^2}\frac{\partial}{\partial\zeta} = \frac{2}{(p^2+1)}(1+\zeta)\frac{\partial}{\partial\zeta},$$

$$\nabla_{\mathbf{p}}\left(\frac{2\mathbf{p}}{(p^2+1)}\frac{\partial}{\partial\xi}\right) = \frac{2}{(p^2+1)}\frac{\partial}{\partial\xi} - \frac{4\mathbf{p}}{(p^2+1)^2}\left(\mathbf{p}\frac{\partial}{\partial\xi}\right),$$

$$\left(\nabla_{\mathbf{p}}\frac{(p^2-1)}{(p^2+1)}\right)\frac{\partial}{\partial\zeta} = \frac{4\mathbf{p}}{(p^2+1)^2}\frac{\partial}{\partial\zeta}..$$

We multiply the first two relations by factor $i\mathbf{p}$ and these next two relations by $-\frac{i(p^2-1)}{2}$. As a result, after summing up given operators the application of the Runge-Lenz operator (24) to the function (27) yields

$$A_{\mathbf{p}} f(\xi,\zeta) = i\left(\xi\frac{\partial}{\partial\zeta} - \zeta\frac{\partial}{\partial\xi}\right) f(\xi,\zeta). \qquad (28)$$

It means that the vector Runge-Lenz operator in the momentum space is tansformed to three infintesimal rotation operators on the Fock sphere, which were previously predicted theoretically. We notice that the commutation properties on the Fock sphere are evident. Additionally, the sum of squared operators (26) is transformed to the angular part of the four-dimensional Laplacian, which has eigenvalue $(n^2 - 1)$ naturally.

## VI. CONCLUSIONS

We find differential equation and the Runge-Lenz operator in the momentum space. They return the quantum Coulomb problem to the area of methods acting in the momentum space. Besides, the Runge-Lenz operator is also useful for interactions involving the SO(4) symmetry. The Fock method is also simplified because an integral equation is no longer necessary and consideration can be performed in the momentum space. Eigenfunctions are as simple as those in the coordinate space and are well integrated for simple pertubations.